\documentclass[a4paper,fleqn]{cas-sc}
\usepackage[hyphenbreaks]{breakurl}
\usepackage{float,url,booktabs,ulem}
\usepackage[longnamesfirst]{natbib}
\usepackage[utf8]{inputenc}
\usepackage[T1]{fontenc}
\usepackage[english]{babel}

\newcommand\lp{\left(}
\newcommand\rp{\right)}
\newcommand\lbr{\left\lbrace}
\newcommand\rbr{\right\rbrace}
\newcommand\lsz{\left|}
\newcommand\rsz{\right|}

\begin{document}

\let\WriteBookmarks\relax
\def\floatpagepagefraction{1}
\def\textpagefraction{.001}

\shorttitle{Topological measures in weighted hypergraphs}

\shortauthors{E. Vasilyeva \textit{et al.}}

\title [mode = title]{Topological measures in weighted hypergraphs}

\author[1,2]{Ekaterina Vasilyeva}\cormark[1]\ead{serebryannikovaee@lebedev.ru}
\author[2,3]{Liubov Tupikina}
\author[2,4]{Daniil Musatov}
\author[2,4,5,6]{Andrei~M. Raigorodskii}
\author[7,8,9]{Charo.~I. del~Genio}
\author[7,8,10]{Stefano Boccaletti}

\address[1]{P.~N. Lebedev Physical Institute of the Russian Academy of Sciences, Leninskiy Prospekt 53, Moscow 119991, Russia}
\address[2]{Moscow Institute of Physics and Technology, Institutskiy Pereulok 9, Dolgoprudny 141701, Russia}
\address[3]{ITMO University, Kronverkskiy Prospekt 49, St.~Petersburg 197101, Russia}
\address[4]{Caucasus Mathematical Center, Adyghe State University, ul.~Pervomayskaya 208, Maykop 385000, Russia}
\address[5]{Faculty of Mechanics and Mathematics, Moscow State University, Leninskie Gory 1, Moscow, 119991 Russia}
\address[6]{Institute of Mathematics and Computer Science, Buryat State University, ul.~Ranzhurova 5, Ulan-Ude 670000, Russia}
\address[7]{Institute of Interdisciplinary Intelligent Science, Ningbo University of Technology, Ningbo, China}
\address[8]{School of Mathematics, North University of China, 030051, Taiyuan, China}
\address[9]{Institute of Smart Agriculture for Safe and Functional Foods and Supplements, Trakia University, Stara Zagora 6000, Bulgaria}
\address[10]{CNR -- Institute of Complex Systems, Via Madonna del Piano 10, I-50019, Sesto Fiorentino, Italy}

\cortext[cor1]{Corresponding author}

\begin{abstract}
Higher-order interactions introduce an additional structural dimension to complex networks,
requiring consistent generalizations of classical topological measures. In hypergraphs, the
definition of distance between nodes is not unique: beyond the conventional measure derived
from clique projection, an alternative formulation that explicitly incorporates the sizes of
hyperedges, those of their intersection and their weights has been recently proposed. Here,
we generalize three distance-based topological measures, namely closeness centrality, betweenness
centrality and node eccentricity, using this new hypergraph distance. Trough tractable illustrative
examples, we demonstrate that the differences between results obtained with the two distances
are systematic and arise from structurally meaningful features of the higher-order networks.
Also, analyzing a series of real-world datasets, we show that hypergraphs can be divided into
three distinct classes, corresponding to the possible dominance of specific orders of interaction
over their general metric structure. This provides practical guidance on the possibility of
limiting the analysis to only some specific interaction orders, reducing its complexity while
maintaining the full information of the system.
\end{abstract}


\maketitle

\section{Introduction}
Characterizing complex networks in terms of some measurable
features is a useful task, as one can use the outcome to perform
a classification of the complex systems that use them as their
supporting structure. This is possible because one can define
macroscopic topological observables whose general behaviour
is common amongst networks of the same type~\cite{newman2003structure, boccaletti2006complex, estrada2012structure, boccaletti2014multi, newman2018networks}.
However, in the last several years, an increasing attention
has been paid to networks with higher-order interactions, whose
structure is not described by classic pairwise graphs, but
rather by hypergraphs~\cite{battiston2020networks,boccaletti2023structure,recon25,nearidsimp25,hypermod,solst25,equi26,highchim26}.
It become thus natural to study how to generalize and extend
the topological measures introduced for traditional networks
to the case of higher-order structures~\cite{aksoy2020hypernetwork,nortier2025higher}.
One of the main peculiarities of this process is that the
definition of such topological quantities may not be unique.

The reason for this is that any such measure ultimately depends on the basic
concept of distance. In traditional networks, the distance between
two nodes is defined as the length of the sequence of  adjacent edges
one needs to traverse to go from one of the two end nodes to the other.
In principle, one could apply the same concept to a higher-order network
by simply replacing edges with hyperedges. However, while in an unweighted
simple graph all edges have size~2, and two adjacent edges only intersect
in one node, the situation is substantially different for
higher-order networks. In fact, in a general hypergraph, the hyperedges
can have any size, and their intersections are not limited to be a
single node. Thus, treating all the hyperedges equally is equivalent
to projecting the hypergraph onto a traditional graph by replacing
every hyperedge with a clique on its constituent nodes. This is indeed a very
common step in the analysis of hypergraphs, even though it is known
to lead to information loss~\cite{boccaletti2023structure}.
To avoid such a drawback, a general distance that is an inherently higher-order
measure has been proposed~\cite{vasilieva}, based on the idea that
the final distance between two nodes should be proportional to the
sizes of the hyperedges that connect them, and inversely proportional
to their intersection. This metric also allows one to specify in a completely
arbitrary way the contribution that hyperedge weights give to the distance,
which is a necessary ability to maintain a reasonable degree of realism
when modelling real-world systems~\cite{vasilyeva2023distances}.

In this article, we generalize the definitions of closeness
centrality, betweenness centrality and
eccentricity to higher-order networks, using the definition
of distance introduced in Ref.~\cite{vasilieva}. We demonstrate
the use of these generalized quantities on a few tractable examples, illustrating the difference
between the values obtained using our method and those that
one gets via clique projection, and explaining their origin.
Also, we compare the results of the two approaches for a number
of real-world higher-order networks, showing that the differences
can be significant. Finally, using the multilayer representation
of hypergraphs~\cite{vasilyeva2021multilayer}, we introduce
the \textit{vector of distances}, where each component corresponds
to a subsequent filtration layer in the hypergraph, showing
that, when studying real-world systems, one can use it to detect
the highest order of interaction that must be included to maintain
an acceptable precision of the results while minimizing the
complexity of the analysis.

\section{Topological observables}
\subsection{Hypergraph distances}
In Ref.~\cite{vasilieva}, an extended concept of distance
for weighted hypergraphs, accounting for hyperedge weights,
hyperedges size and different hyperedge intersection, was
introduced. This new definition was built
to satisfy the following properties: (1) the distance between
two nodes that participate in the same hyperedge must increase
with hyperedge size; (2) the distance between two nodes belonging
to different hyperedges must decreasing with the size of
the intersection of the hyperedges; (3) it must be possible
to tune the dependence of the distance from the hyperedge
weights to reflect the nature of the weights. Additionally,
the measure must be well-defined and reduce to the classical
graph distance measure when all the hyperedges are of size~2.
Note that the specific form proposed in Ref.~\cite{vasilieva}
is not necessarily the only one that meets these requirements,
even though it is shown to be a reasonable choice. For the
sake of self-consistency, we briefly recall here the relevant
definitions.

To define the new hypergraph distance,
given a weighted hypergraph~$H^w$, one first constructs
its weighted line graph with loops~$L(H^w)$. To do so,
one creates a node in~$L(H^w)$ for each hyperedge of~$H^w$,
and, given two nodes~$e$ and~$\tilde e$ of~$L(H^w)$, one
creates an edge between them if and only if~$e$ and~$\tilde e$
have at least one node in common within~$H^w$. The weights
of the edges of~$L(H^w)$ are given by
\begin{equation}\label{linegraph-w'}
 w'_{e\tilde e} =\frac{1}{2} \lp f(W_e)\left| e\right| + f(W_{\tilde e})\left| \tilde e\right| + \frac{f(W_e)\left| e\right| + f(W_{\tilde e})\left|\tilde e\right|}{\left|e \cap\tilde e\right|}\rp -3\frac{f(W_e)+f(W_{\tilde e})}{2},
\end{equation}
where~$W_e$ is the weight of hyperedge~$e$ and~$f$
is any positive function that encodes a relationship
between the hyperedge weights and the concept of distance. The specific
form of Eq.\eqref{linegraph-w'} is such that the weights of the edges of the line graph
satisfy the required properties of the distance measure described above.

Having constructed the line graph, the distance between
two nodes~$i$ and~$j$ in~$H^w$ is defined as
\begin{equation}\label{dw}
 d^w(i,j) = \begin{cases}
  \displaystyle{\min_{\substack{e\in E_i\\\tilde e\in E_j\\{}}}}\lp d^{L'}(e, \tilde e) + f(W_e)\left| e\right| + f(W_{\tilde e})\left|\tilde e\right|-3\frac{f(W_e)+f(W_{\tilde e})}{2}\rp &\quad\text{if $i\neq j$,}\\
  0 &\quad\text{if $i=j$,}
  \end{cases}
\end{equation}
where~$d^{L}(e,\tilde e)$ is the classic
graph distance between~$e$ and~$\tilde e$
in~$L(H^w)$.

Also, in the following we indicate with~$d^p(i,j)$ the distance between
node~$i$ and node~$j$ computed via the weighted clique
projection~$G^w$ of~$H^w$. We recall that~$G^w$ is a weighted
undirected graph on the same nodes as~$H^w$, and where
an edge between two nodes exists if they are connected
by at least one hyperedge in~$H^w$. The edge weights in~$G^w$
are a function of the number of hyperedges that connect
the corresponding nodes in~$H^w$. Below, when discussing
the differences between the results obtained using~$d^w(i,j)$
and those obtained from~$d^p(i,j)$, we use the same function~$f$
in both formulations, to guarantee the sensibility of
the comparison. Specifically, for a given hyperedge weight~$x$,
we put $f=\frac{1}{x}$. This is a natural choice
for cases in which higher weight of hyperedges represent closer connections
between the nodes that constitute them. In particular,
it is consistent with the nature of weights in the real-world
examples considered in Section~\ref{data}, which relate
to the concept of cognitive distance. Finally, we will henceforth refer to~$d^w(i,j)$
as the \textit{hypergraph distance} and to~$d^p(i,j)$ as
the \textit{projected distance}.

\subsection{Distance-based observables}
The idea of distance is used in graph theory
as a starting point to introduce several topological
observables that quantify specific properties
of a complex network. Here, we focus on how
one can generalize closeness centrality, betweenness
centrality and node
eccentricity to hypergraphs, and how such an
extension provides different levels of information
depending on the definition of distance adopted.

Closeness centrality is one of the most frequently
studied centrality measures, used to rank the nodes
in a way that is inversely proportional to their distance
from all other nodes~\cite{newman2003structure,boccaletti2006complex,estrada2012structure}.
In formulae, the closeness centrality~$C_i$ of node~$i$
in a network of $N$~nodes is
\begin{equation}\label{clos}
 C_i = \frac{1}{\displaystyle\sum_{\substack{j=1\\j\neq i}}^N d(i,j)}\:,
\end{equation}
where~$d(i,j)$ is the classic graph distance.
Closeness centrality is readily generalized to
higher-order structures by replacing~$d(i,j)$
with the formulation of distance in hypergraphs
one decides to use. Below, we consider $C^p_i$,
which uses the projected distance, and~$C^w_i$,
which uses the hypergraph distance.

\begin{figure}[t]
 \centering
 \includegraphics[width=0.5\textwidth]{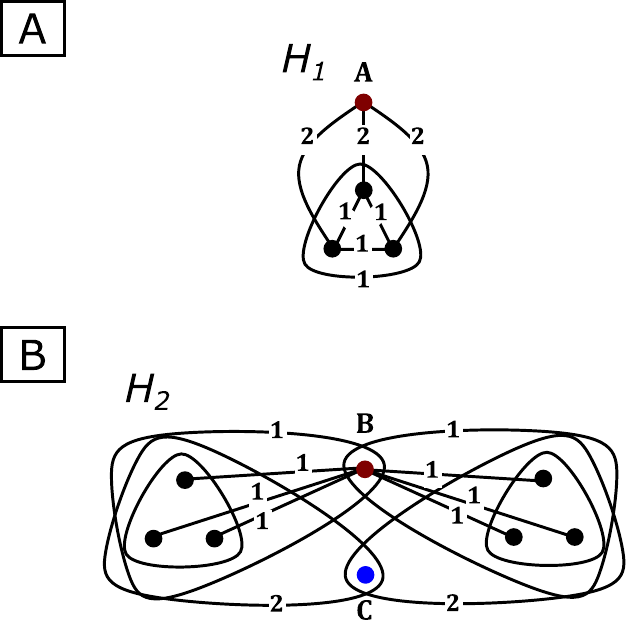}
 \caption{\textbf{Illustrative examples of hypergraphs.}
 A) Hypergraph~$H_1$ contains four nodes connected by
 7~edges, one of size~3 and the rest of size~2. B) Hypergraph~$H_2$
 contains 8~nodes and 12~edges. Nodes~B and~C act as bridges
 between two sets of densely connected nodes.}\label{fig:ex}
\end{figure}
Another fundamental distance-based centrality measure
is betweenness centrality, which measures the importance
of a node in transport processes on the network.
Classically, the betweenness centrality $b_i$ of node~$i$
is given by
\begin{equation}\label{betw}
 b_i = \sum_{\substack{u,v=1\\v>u\\u,v\neq i}}^N\frac{\left|P_i(u,v)\right|}{\left|P(u,v)\right|}\:,
\end{equation}
where $P(u,v)$ is the set of all shortest paths
between~$u$ and~$v$, and $P_i(u,v)$ is the set
of shortest paths between~$u$ and~$v$ that pass
through~$i$. To generalize this measure to the
case of hypergraphs we follow the procedure described
in~\cite{vasilyeva2023distances,puzis2013betweenness}.
The key idea is that one cannot replace directly
the quantities in the equation above with their
hypergraph equivalents, because while
in a traditional graph the size of the intersection
of two adjacent edges is always~1, this is not
necessarily true in hypergraphs, where such an intersection
can contain more than one~node. Thus, since a path
$\pi=\lbr e_1,e_2,\dotsc,e_\ell\rbr$ is technically
defined as sets of (hyper)edges, explicitly accounting
for the size of their intersections is necessary
to maintain the relation between betweenness and
connectivity. Therefore, we define the betweenness centrality
of a node~$i$ in a hypergraph as
\begin{equation}
 b_i = \sum_{\substack{u,v=1\\v>u\\u,v\neq i}}\frac{\displaystyle\sum_{\pi\in P_i(u,v)}\frac{1}{\displaystyle\sum_{j=1}^{\ell-1}\left|e_j\cap e_{j+1}\right|\mathbf 1_{i\in\lbr e_j\cap e_{j+1}\rbr}}}{\left|P(u,v)\right|}\:,
\end{equation}
where~$\mathbf 1$ is an indicator variable.
In other words, we don't just count the number
of shortest paths between two nodes that pass
through~$i$, but, for each such path, we check
all the pairs of consecutive hyperedges that
have~$i$ in their intersection, and sum the
inverses of the sizes of such intersections.
To distinguish between the betweennesses obtained
by the two different definitions of distance,
we will indicate with~$b^p_i$ the betweenness
resulting from the use of the projected distance,
and with~$b^w_i$ the one obtained with the
use of the hypergraph distance.

\begin{figure}[t]
 \centering
 \includegraphics[width=0.9\textwidth]{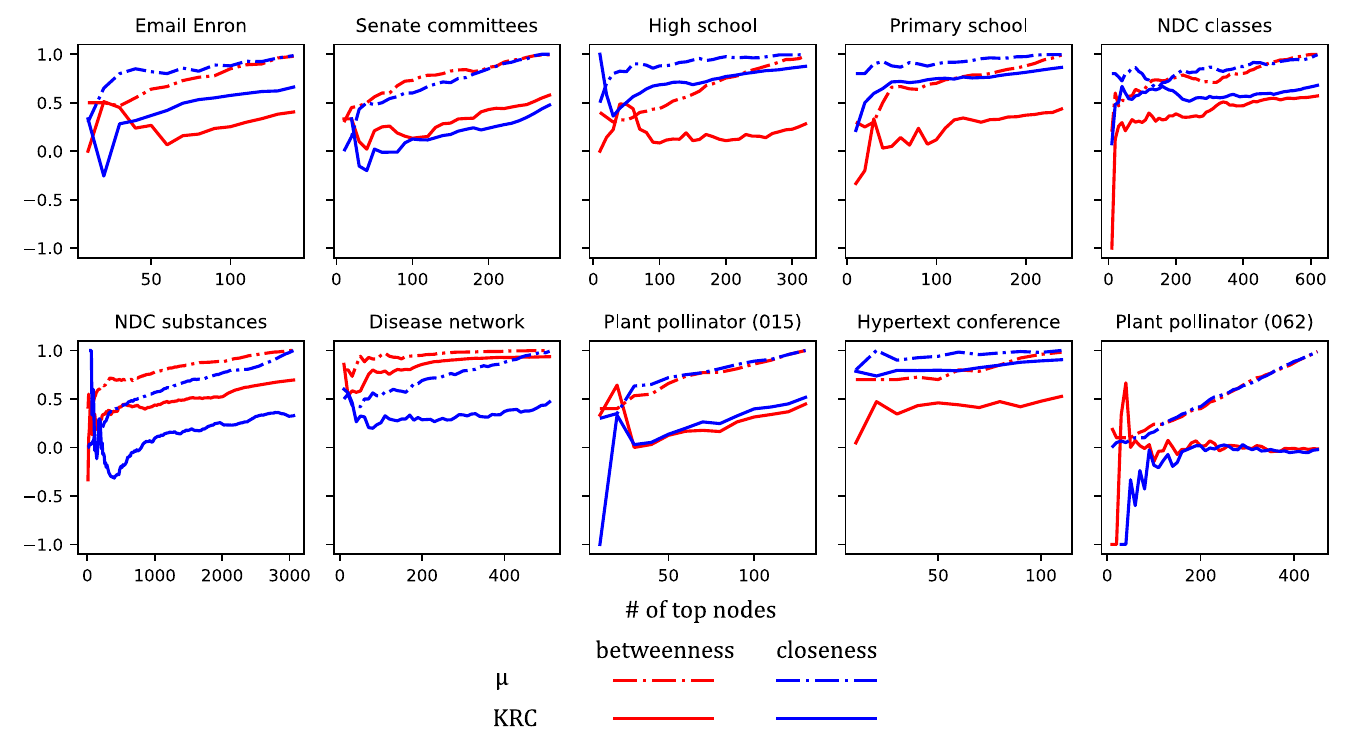}
 \caption{\textbf{Similarity of node rankings by closeness
 and betweenness.} The behaviours of the~$\mu$-measure
 and of the Kendall rank correlation coefficient~(KRC)
 with the number of included top nodes reveals how the
 importance of higher-order connections changes at different
 scales. In particular, they are more relevant at the
 global scale than at the local one in the school dataset
 and in the conference contact network, whereas the situation
 is the opposite for the disease network. For all the
 other datasets the higher-order connections dominate
 at both scales.}\label{centr_gen}
\end{figure}
Finally, we consider the eccentricities of the nodes.
Eccentricity is normally used to define the so-called center
of a graph, which is a particularly important concept
when the graph is a tree. By definition,
the eccentricity $\varepsilon_i$ of a node~$i$ is
the maximum distance between~$i$ and any other node
in the graph. Then, the center of the graph is defined
as the set of nodes with the lowest eccentricity,
and one can easily prove that in trees the center
consists either of a single node or of two adjacent
nodes. This quantity can be easily generalized to
hypergraphs, since its definition remains
\begin{equation}
 \varepsilon_i = \max_{v\in V}d(i,v)\:,
\end{equation}
where $V$ is the set of nodes of the hypergraph.
Note that, while eccentricity
in traditional graphs is an integer, the definitions
we use for it in hypergraphs make it, in general,
a real-valued observable. Also, consistently with
the formalism adopted so far, we will call~$\varepsilon_i^p$
and~$\varepsilon_i^w$ the eccentricity of node~$i$
in a hypergraph computed using projected distance
and hypergraph distance, respectively.

To estimate the computational complexity
of computing the hypergraph observables considered here,
first note that, in the case of closeness and eccentricity,
it is directly related to finding the distances between
all node pairs, and in the case of betweenness centrality,
it is related to finding all shortest paths. However, the
time complexity for these problems is asymptotically the
same. Therefore, the dominant contribution can be estimated
from the same considerations as in Ref.~\cite{vasilieva}.
Ultimately, this yields an average complexity of $O\left(M\left( D+M\log M\right)+N^2{\bar d}^2\right)$,
where $M$ is the number of hyperedges, $D$ is the number
of intersecting hyperedges, $N$ is the number of nodes and
$\bar{d}$ is the average degree of the nodes in the hypergraph.
In turn, this implies a worst-case complexity of $O(M^3)$.

To better study the structure of interactions
of real-world higher-order networks, we also
introduce the idea of \textit{distance vector}.
To do so, we start from the multilayer representation
of a hypergraph~$H^w$~\cite{vasilyeva2021multilayer}.
If $K$ is the maximum size of the hyperedges
found in~$H^w$, then for all values of~$k$ between~2
and~$K$, we define~$H^{w,k}$ as the sub-hypergraph
of~$H^w$ containing all hyperedges whose size
is not larger than~$k$. This sub-hypergraph
also takes the name of $k$-th layer of~$H^w$.
We can then indicate with~$d^{w,k}(i,j)$ the
distance between node~$i$ and node~$j$ in~$H^{w,k}$
computed using Eq.~\ref{dw}, and, based on this,
we introduce the distance vector $\mathbf{d^w}(i,j)$
between two nodes~$i$ and~$j$ in~$H^w$ as the
vector of size~$K-1$ given by
\begin{equation}
 \mathbf{d^w}(i,j) = \lp d^{w,2}(i,j), d^{w,3}(i,j), \dotsc, d^{w,K}(i,j)\rp^{\mathrm T}\:.
\end{equation}
Crucially that the components~$d^{w,k}(i,j)$
of~$\mathbf{d^w}(i,j)$ are defined only if~$i$ and~$j$ belong
to the same component in layer~$k$, i.e., in the sub-hypergraph~$H^{w,k}$.
Otherwise, if a given pair of nodes~$i$ and~$j$ first belong
to the same component in layer~$k$, then the first $k-2$~components
of~$\mathbf{d^w}(i,j)$ are undefined, and the properties of
distance evolution are only considered from layer~$k$ onwards.

Note that one can also define a distance vector
using clique projection, whose elements are the
distances $d^{p,k}(i,j)$ in the weighted clique
projection of~$H^{w,k}$. Also, being defined on
weighted hypergraphs, all these quantities naturally
include the case of unweighted hypergraphs.

\section{Data}
\label{data}
\begin{figure}[t]
 \centering
 \includegraphics[width=0.9\textwidth]{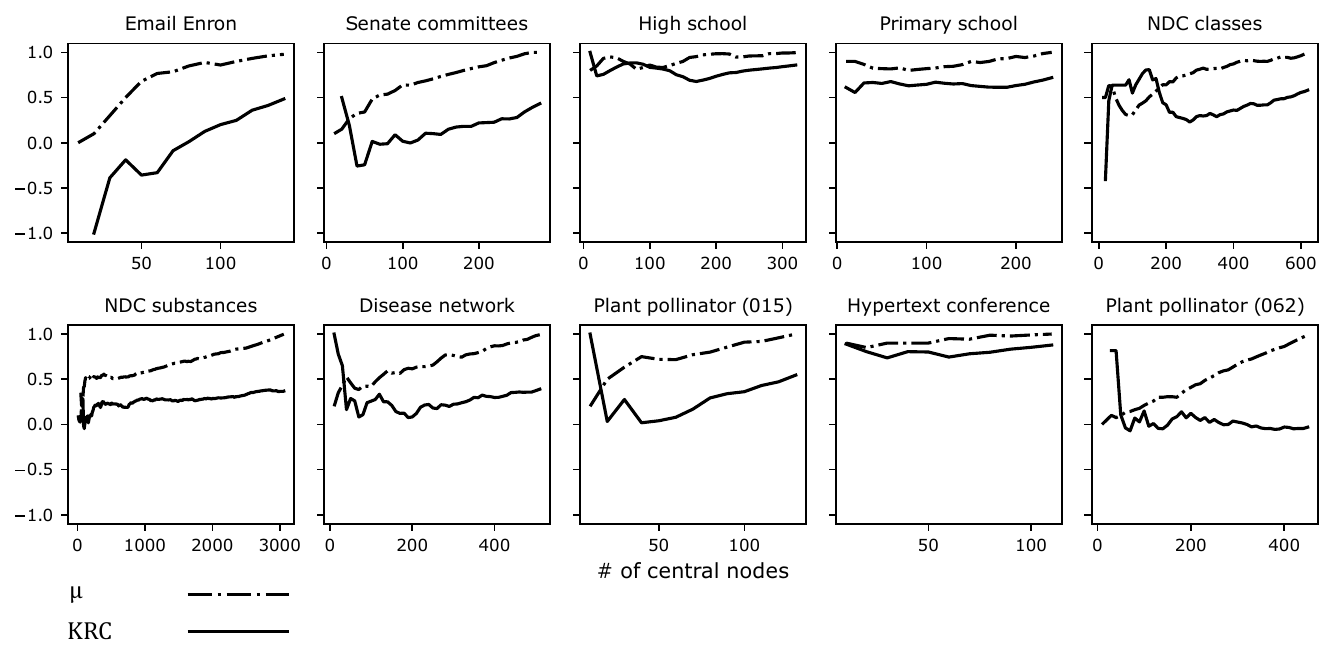}
 \caption{\textbf{Similarity of node rankings by eccentricity.}
 The behaviours of the~$\mu$-measure and of the Kendall rank
 correlation coefficient~(KRC) with the number of included top
 nodes reveals that, if higher-order interactions are not negligible,
 the topological structure of the networks changes radically
 with the chosen definition of distance. Thus, while the differences
 are small in the contact networks (primary school, high school
 and hypertext conference), they become very large in all the
 other cases, and especially for the Enron email network and
 the plant pollinator~(062) one.}\label{eccentr_gen}
\end{figure}
\begin{table}[t]
\begin{tabular}{@{}cccccr@{.}lccr@{.}lc@{}}
\toprule
Dataset & $\lsz V\rsz$ & $\lsz E\rsz$ & $\max\lsz e\rsz$ & $\min\lsz e\rsz$ & \multicolumn{2}{c}{Mean $\lsz e\rsz$} & Median $\lsz e\rsz$ & $\max W_e$ & \multicolumn{2}{c}{Mean $W_e$} & Median $W_e$\\ \midrule
High school       &  327 &  7818 &   5 & 2 &  2 & 3 &  2 & 2772 & 22 & 0 & 2 \\
Primary school    &  242 & 12704 &   5 & 2 &  2 & 4 &  2 &  601 &  8 & 4 & 2 \\
Enron emails      &  143 &  1457 &  18 & 2 &  3 & 1 &  2 &  241 &  7 & 2 & 2 \\
NDC substances    & 3065 &  6093 &  25 & 2 &  8 & 1 &  6 & 1296 &  4 & 6 & 1 \\
NDC classes       &  628 &   794 &  24 & 2 &  7 & 1 &  5 & 2083 & 50 & 0 & 5 \\
Senate committees &  282 &   301 &  31 & 4 & 17 & 6 & 19 &    3 &  1 & 0 & 1 \\
Diseases          &  516 &   314 &  11 & 2 &  3 & 0 &  3 &    4 &  1 & 1 & 1 \\
Hypertext conf.   &  113 &  2434 &   6 & 2 &  2 & 2 &  2 & 1252 &  7 & 8 & 2 \\
Pollinators 015   &  130 &   401 & 104 & 2 &  6 & 6 &  4 &    4 &  1 & 0 & 1 \\
Pollinators 062   &  456 &   866 & 157 & 2 & 17 & 4 & 10 &    1 &  1 & 0 & 1 \\ \bottomrule
\end{tabular}\caption{\textbf{Descriptive statistics of the benchmark datasets.}
The columns report, in order, number of nodes, number of hyperedges, maximum hyperedge
size, minimum hyperedge size, mean hyperedge size, median hyperedge size, maximum
hyperedge weight, mean hyperedge weight and median hyperedge weight. The minimum
hyperedge weight is~1 for all the datasets.}\label{tab:stats}
\end{table}
In this study, we use a number of real-world datasets describing
systems with higher-order interations of diverse nature. Specifically,
we analyze 10~real-world datasets that are commonly used as a benchmark
for the analysis of hypergraph properties, and, in addition
we study hypergraphs constructed from the metadata of the preprints
published on the arXiv. While choosing the list of datasets,
we were guided by the following considerations. First, we decided
not to consider datasets that were trivially small, while at the
same time excluding those that were excessively large. Second, we
chose datasets that refer to broad subject fields. Thus, for example,
from ten different plant-pollinator datasets that are available in the
XGI~repository, we selected two with the largest number of nodes.
Below, we provide a brief description of
each dataset. The basic topological properties of the first 10~datasets
are reported in Table~\ref{tab:stats}, whereas those of the arXiv
hypergraphs can be found in~\cite{vasilieva}.
\begin{itemize}
 \item \textbf{High-school/Primary-school contacts} \cite{Benson-2018-simplicial, Mastrandrea-2015-contact}:
 in these two datasets, nodes represent students enrolled in a primary school or in a high school, and hyperedges
 represent contacts between them. Note that hyperedges are not unique and their frequency can be interpreted as the
 duration of the contact.
 \item \textbf{Enron emails} \cite{Benson-2018-simplicial}: nodes represent email addresses of the Enron company
 employees and a hyperedge connects the sender and all recipients of each email. The multiplicity of a hyperedge
 corresponds to the number of exchanged emails with the same set of people involved.
 \item \textbf{NDC substances/classes} \cite{Benson-2018-simplicial}: each hyperedge in these two datasets
 corresponds to the National Drug Code~(NDC) for a drug, and the nodes within each hyperedge are the substance/class
 labels of the components. Note that these hypergraphs are not connected; thus, whenever we refer to them below,
 we actually intend their largest connected components.
 \item \textbf{Senate committees} \cite{chodrow2021generative}: nodes represent members of the US~Senate
 and hyperedges correspond to committee memberships. In this dataset 288~hyperedges out of a total of~301 are unique.
 \item \textbf{Disease network} \cite{goh2007human}: this dataset links diseases and the genes associated with them.
 Specifically, nodes correspond to diseases and hyperedges to the implicated genes.
 \item \textbf{Plant pollinator networks 015 and 062} \cite{petanidou1993pollination,robertson1929flowers}:
 in these datasets, nodes are plant species and hyperedges are pollinator species that visit a given plant.
 The data of dataset~015 were collected in Daphni, Athens, Greece, whereas those of dataset~062 were collected
 in Carlinville, IL, USA.
 \item \textbf{Hypertext conference} \cite{isella2011s}: this dataset contains information on active contacts
 between participants to the ACM~Hypertext~2009 conference, recorded over 20-second intervals.
 \item \textbf{arXiv} \cite{clement2019arxiv}: the arXiv dataset contains scientific preprints uploaded
 online between~1992 and~2018. In this work we use a series of 1-year sub-samples relating to the period
 1998--2018. Each preprint is marked by series of tags representing scientific fields to which it relates.
 Therefore, in the hypergraphs we construct, nodes are represented by scientific-field tags, hyperedges represent
 preprints having specific subsets of tags, and the weights of the hyperedges are their frequencies of
 occurrence.
\end{itemize}

\section{Results}
\subsection{Analysis of the distance-based topology measures}
\subsubsection{Illustrative examples}
To emphasize the usefulness of the new measures in the description
of higher-order topologies, we first describe a series of illustrative
tractable examples, shown in Fig.~\ref{fig:ex}. These demonstrate
that the differences between the results obtained using hypergraph
distance and projected distance actually arise from additional information
about the higher-order structure that is lost when projecting but
explicitly accounted for by the hypergraph distance. Note that, while
being relatively simple to maximize clarity, the examples we give
are small structures that frequently arise as motifs in large real-world
hypergraphs.

The first example, hypergraph~$H_1$ visualized in Fig.~\ref{fig:ex}(A),
is useful to illustrate the effect of the distance definitions on closeness
centrality and eccentricity. Start by considering the closenesses of node~A,
$C^w_A$ and~$C^p_A$, and those of the other nodes, $C^w_i$ and~$C^p_i$.
By carrying out an explicit calculation, we find that the closeness of
node~A is~$0.67$, regardless of which distance definition one uses. However,
the closeness of the other nodes is also~$0.67$ when using the projected
distance, but it decreases to~$0.4$ with the hypergraph distance. This
follows from the fact that the weighted clique projection of~$H_1$ is
just a clique with all edge weights equal to~2. Thus, the projection completely
flattens the information about the different types of edges, even though
node~A is connected to the rest of the nodes more strongly than they are
amongst themselves. One can imagine a similar situations in scientific
collaboration networks, with one author (node~A) writing 2~articles with
each of three other authors, while these other authors have one additional
joint article and three more papers in pairs. In such a structure we would
expect node~A to be somehow more central than the other nodes in terms
of closeness and, indeed, this is indeed the case when comparing~$C^w_A$ and~$C^p_A$. Thus, the new measure assesses node~A more realistically than
the projected one. A similar effect is found on the eccentricity, where
the projected one is~$0.5$ for all nodes, whereas the hypergraph one is~$0.5$
for node~A and~1 for all other nodes.

\begin{figure}[t]
 \centering
 \includegraphics[width=0.9\textwidth]{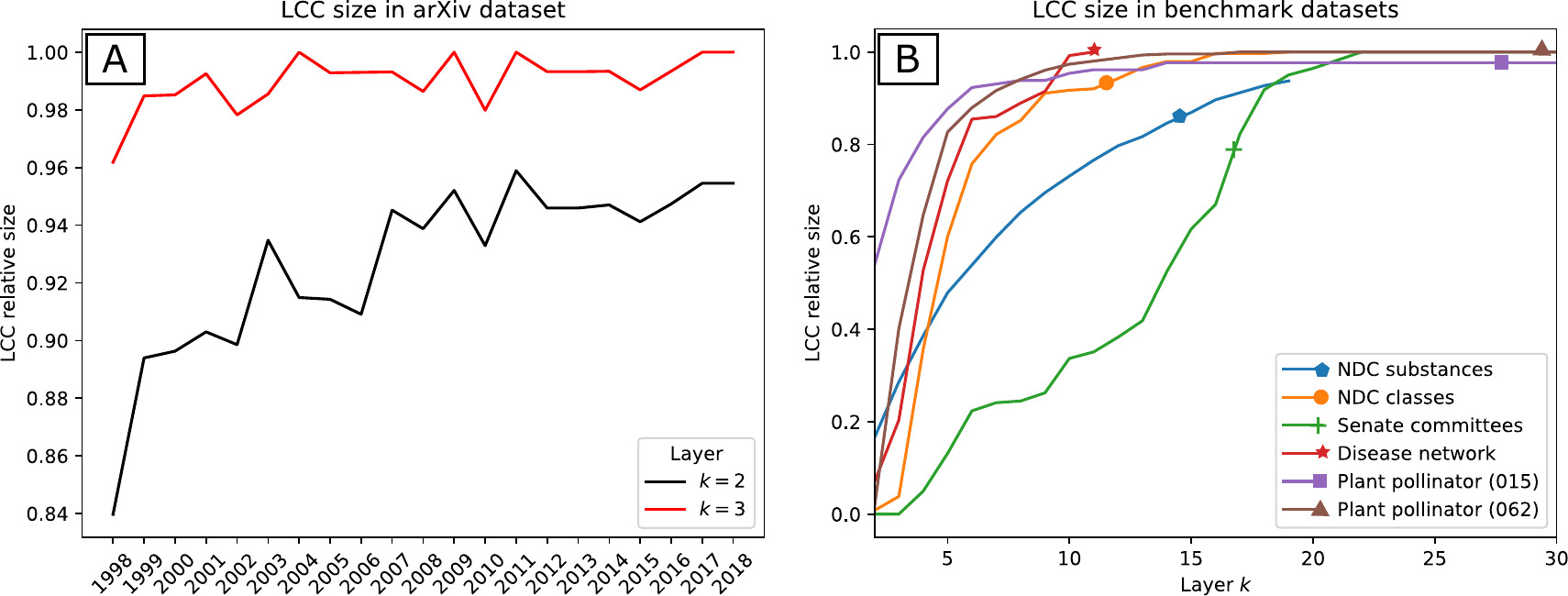}
 \caption{\textbf{Relative size the of the largest
 connected component (LCC) in hypergraph layers.} A)
 In the arXiv dataset, almost all nodes are connected
 already when only pairwise interactions are considered
 ($k=2$, black curve). For $k=3$, all the nodes belong
 to the LCC in some years. B) In the real-world datasets,
 the fractions of nodes in the LCC increase differently
 with the layers considered, with the slower increases
 corresponding to networks with more prominent higher-order
 interactions. In all cases the sizes of the LCC are
 relative to the LCC size in the whole hypergraph.}\label{lcc}
\end{figure}
As second example, consider hypergraph~$H_2$ in Fig.~\ref{fig:ex}(B),
which provides a good means to illustrate differences in betweenness
centrality. Since betweenness identifies important nodes for the spreading
processes happening along the shortest path, we focus on nodes~B and~C,
which act as a pair of bridges between two sets of three nodes that are
densely connected. Here, the difference between projected measure and
hypergraph measure is even more dramatic than before. In fact, the projected
betweenness is~$4.5$ for both~B and~C. However, the hypergraph betweenness
is~15 for node~B, but~0 for node~C, indicating that, when the higher-order
nature of the edges is accounted for, not one shortest path passes through~C.
This is easily explained by noting that node~B also forms pairwise edges
with all other nodes of the hypergraph except~C, and that the hyperedges
involving~B and other nodes have weight~1, whereas those involving~C
and the other nodes have weight~2.

\subsubsection{Real-world cases}
Having illustrated how the hypergraph measures yield
results that reflect the structure of higher-order connections
within the networks, we apply them to the real-world
datasets, described above.

\paragraph{Betweenness and closeness}
First, we analyze the differences between projected distance
and betweenness, and their hypergraph counterparts. Similarly
to the approach used previously in Ref.~\cite{vasilyeva2023distances,kovalenko2022vector},
we measure the similarity of the rankings of the top~$t$ nodes
with respect to pairs of projected and hypergraph centralities.
Specifically, given two rankings~$\mathbf x$ and~$\mathbf y$,
we quantify their similarity by means of the Kendall rank correlation
coefficient~(KRC) and of the $\mu$-measure, defined as
\begin{equation}\label{mu}
 \mu_t (\mathbf x, \mathbf y) = \frac{\left|T_t(\mathbf{x})\cap T_t(\mathbf{y})\right|}{t}\:,
\end{equation}
where~$T_t(\mathbf x)$ is the set of the $t$~top nodes in the ranking~$\textbf{x}$.

\begin{figure}[t]
 \centering
 \includegraphics[width=0.9\textwidth]{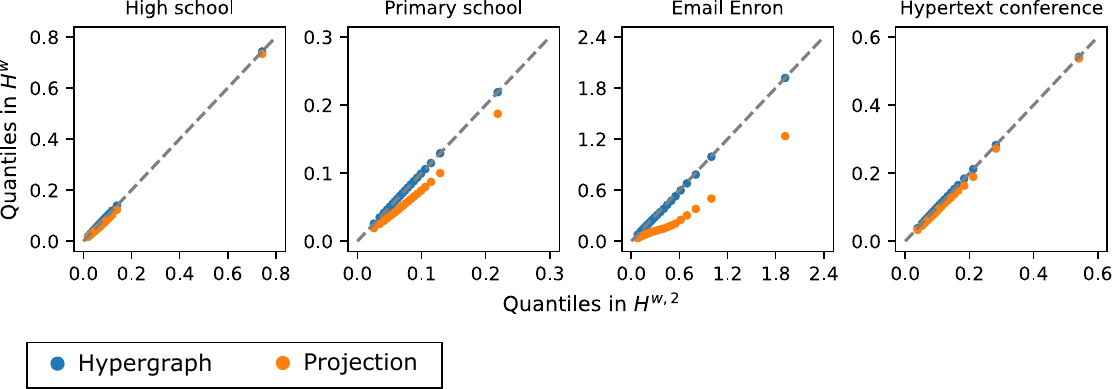}
 \caption{\textbf{Quantile-quantile distance plots for benchmark datasets
 with pairwise interaction dominance.} The panels show the correspondence
 between the quantiles of the distributions of distances $d^{w,2}(i,j)$
 and $d^w(i,j)$ (blue), and those of the distributions of distances $d^{p,2}(i,j)$
 and $d^{p}(i,j)$ (orange). The hypergraph distances are dominated by pairwise
 interactions, as it is evident from the distance remaining unchanged upon
 the addition of all the layers.}\label{pair_dominance_bench}
\end{figure}
The results of betweenness and closeness,
illustrated in Fig.~\ref{centr_gen}, show
similarities and differences between groups
of datasets. In particular, both the projected
and the hypergraph measures yield fairly
similar sets of top nodes for both closeness
and betweenness in the primary-school, high-school
and hypertext-conference datasets, for which
the nature of the interactions is that of
personal contacts. This is particularly evident
in the conference dataset, where also the
rankings according to closeness are very
similar. The rankings according to betweenness,
however, exhibit marked differences in all
three datasets. The same qualitative behaviour
is found, in a less pronounced way, in the
school data for closeness rankings, where,
additionally, the sets of top nodes for betweenness
become very similar only when a substantial
fraction of nodes is considered. This indicates
that, while clique projection captures the
local structure of these datasets reasonably
well, it fails to detect the global importance
of specific nodes, providing a different
ranking of the important bridges for information
transfer.

\begin{figure}[t]
 \centering
 \includegraphics[width=0.7\textwidth]{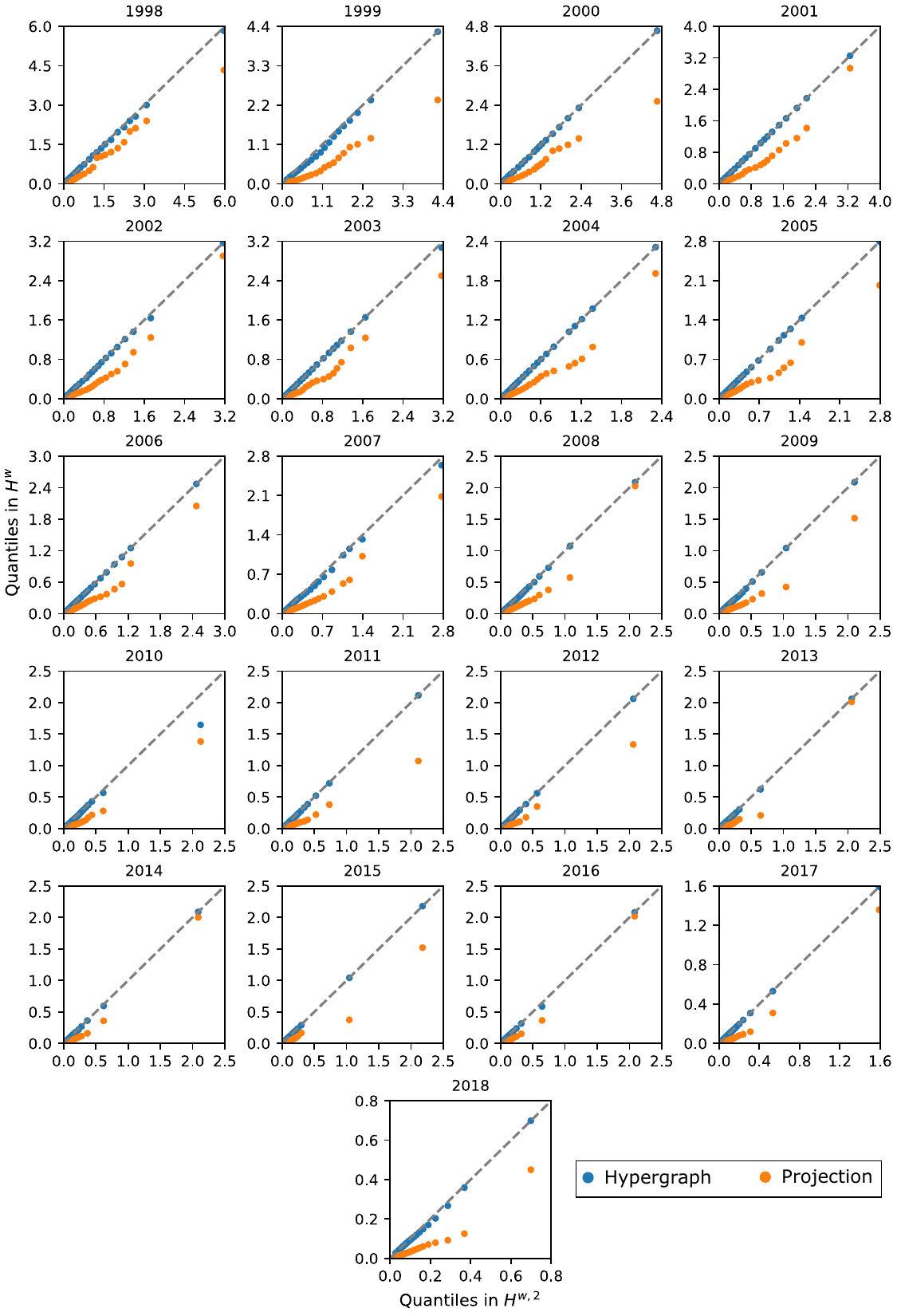}
 \caption{\textbf{The quantile-quantile distance plots for the arXiv
 datasets show a strong pairwise interaction dominance.} The panels
 show the correspondence between the quantiles of the distributions
 of distances $d^{w,2}(i,j)$ and $d^w(i,j)$ (blue), and those of the
 distributions of distances $d^{p,2}(i,j)$ and $d^{p}(i,j)$ (orange).
 Since the hypergraph distances do not change when adding layers to
 the network, the metric structure of these datasets is dominated by
 pairwise interactions.}\label{pair_dominance_arxiv}
\end{figure}
The opposite trend is found for the disease network,
where the most central nodes in terms of betweenness
remain the same, but closeness ranking exhibits marked
differences when comparing the results based on projected
distance with those based on hypergraph distance. This
suggests that the importance of higher-order edges is
more pronounced at a local scale, so that the shortest
paths pass preferentially through the same nodes regardless
of the type of distance used, and even though the distances
between nodes themselves change.

Finally, the remaining datasets yield profound
differences for both centralities. These are especially
dramatic for the plant-pollinator networks, where
the projected and hypergraph rankings are actually
anticorrelated, either only initially (for the
015~dataset) or for any number of top nodes included
(for the 062~dataset). Consistently, for both
datasets, the sets of top nodes are substantially
different. In all these cases, we can conclude
that the higher-order nature of the connections
is a fundamental feature of the systems, and thus
clique projection leads to significant loss of
information.

\paragraph{Eccentricity}
The comparison of the results for eccentricity, shown in Fig.~\ref{eccentr_gen},
paint a picture that complements the previous results with some additional
insight. First, the contact data, i.e, the school datasets and the conference
one, show only small differences in node rankings and top-node sets between
projected eccentricity and hypergraph one. All together, this indicates that
these datasets are dominated by pairwise interactions, which is a reasonable
consideration given their nature. However, all other datasets result in very
strong differences between the two measures. Since the eccentricity measures
the largest distance of any node from a chosen point in the networks, this
shows that, as soon as higher-order interactions become more relevant, the
topology of the network is radically affected by the choice of distance definition.
In fact, while an effect of the metric on the structure is certainly to be
expected, we note that its extent is such that some rankings are even anticorrelated,
as it happens for the Enron network, or completely uncorrelated even in the
top nodes, as in the pollinator~(062) network.

\subsection{Analysis of the multi-layer hypergraphs representation}
\subsubsection{Evolution of connectivity}
The distribution of distances between node pairs
is naturally defined only for the nodes belonging
to the same connected component. Therefore, before
analyzing it, we first measure the evolution of
the largest connected component~(LCC) with the layer
of the hypergraph in the multi-layer representation.
Note that this property does not depend on the type
of distance used.

In the case of the arXiv hypergraphs, most nodes
are connected even when if we restrict ourselves
to only pairwise interactions, as seen in Fig.~\ref{lcc}(A).
In fact, the fraction of nodes in the LCC when considering
layer~2 is always approximately 90\%, rising to~96\%
or more when including also triadic interactions.

The trend is markedly different for the benchmark datasets,
where different behaviours are observed. In particular, for
school data and the conference ones, pairwise interactions
connect all the nodes in the datasets. A similar situation
occurs for the email data, where the LCC size for $k=2$ is
more than 99\%. However, as shown in Fig.~\ref{lcc}(B), the
size of the LCC in the different layers of the hypergraph relative
to the one in the full network increases more slowly for the
other datasets, providing evidence of the higher importance
of higher-order interaction in such cases.

\subsubsection{Classification of the observed vector distances patterns}
\begin{figure}[t]
 \centering
 \includegraphics[width=0.9\textwidth]{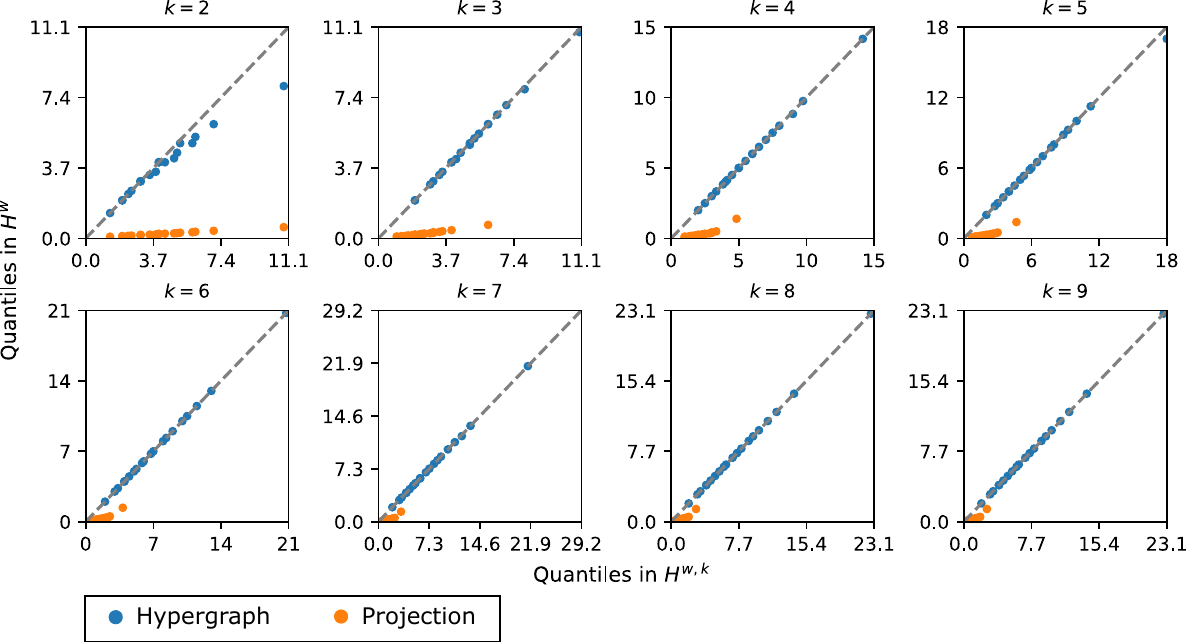}
 \caption{\textbf{The quantile-quantile distance plots for the
 Plant Pollinators~(015) dataset show the dominance of intermediate
 orders of interaction.} The panels show the correspondence between
 the quantiles of the distributions of the distances $d^{w,k}(i,j)$
 and $d^w(i,j)$ (blue), and those of the distributions of the
 distances $d^{p,k}(i,j)$ and $d^{p}(i,j)$ (orange), as more
 layers~$k$ are progressively considered. The two distributions
 of hypergraph distances converge for $k=6$, indicating that
 intermediate-order interactions are the most important in this
 dataset.}\label{lower_dominance}
\end{figure}
\begin{figure}[t]
 \centering
 \includegraphics[width=0.75\textwidth]{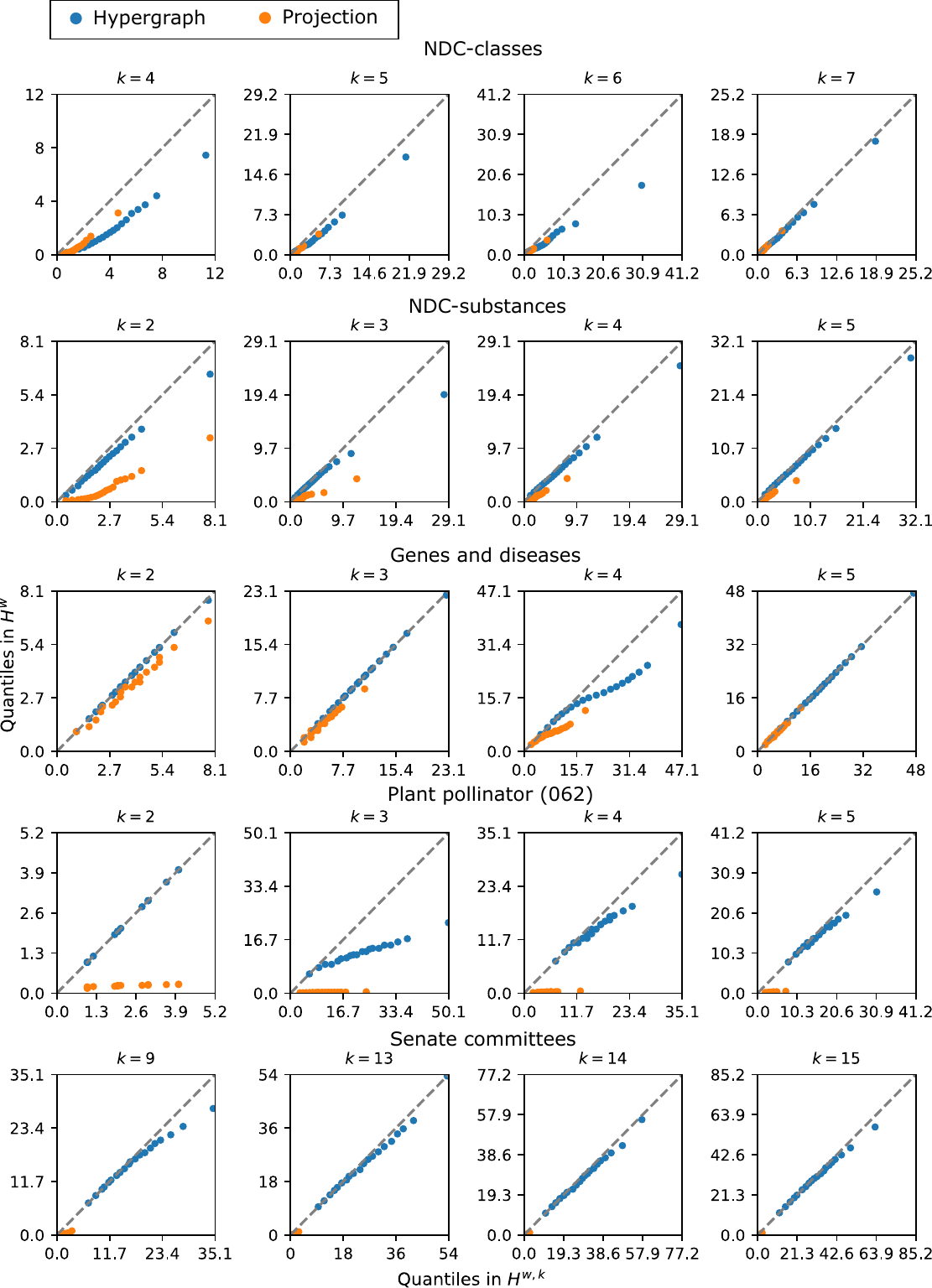}
 \caption{\textbf{The quantile-quantile distance plots for
 the show the dominance of higher-order interactions.} The
 panels show the correspondence between the quantiles of
 the distributions of the distances $d^{w,k}(i,j)$ and $d^w(i,j)$
 (blue), and those of the distributions of the distances
 $d^{p,k}(i,j)$ and $d^{p}(i,j)$ (orange) for datasets in
 Class~3, as more layers~$k$ are progressively considered.
 For all these datasets, the distributions converge only
 when all the interaction orders are considered.}\label{no_dominance}
\end{figure}
The study of how the distances between pairs of nodes change
as one increases the number of layers considered provides information
about the balance between different orders of interaction in
the network. Thus, based on the properties of the distance vector~$\mathbf{d^w}$,
we can assign a hypergraph to one of three broad classes:
\begin{enumerate}
 \item \textbf{Pairwise interaction dominance}.
 In networks belonging to this class, almost all
 the nodes in the LCC are connected for $k=2$.
 Moreover, their distances do not change with the
 addition of more layers. Thus, such networks are
 characterized by a very strong pairwise backbone,
 which dominates their metric structure.
 \item \textbf{Low-order interaction dominance}.
 For networks in this class, the size of the LCC
 grows with the addition of new layers. Thus, there
 are several node pairs for which the distance
 can only be computed starting from some given
 layer $k>2$. Nonetheless, once the distance between
 two nodes can be computed, it remains unchanged
 when new layers are added.
 \item \textbf{High-order interaction dominance}.
 For hypergraphs in this group, the addition of
 higher-order layers leads to a significant decrease
 in the distances between nodes that were already
 connected in lower layers. Thus, to account for
 the full information of the higher-order topology,
 one must explicitly consider all interaction sizes.
\end{enumerate}

In relation to the dataset we analyzed, we find that Class~1
is typical for the contact datasets, such as the school ones
and the Hypertext Conference, even though also the Enron email
network and the arXiv data show a strong dominance of pairwise
interactions. In fact, their quantile-quantile distance relations,
reported in Fig.~\ref{pair_dominance_bench} and Fig.~\ref{pair_dominance_arxiv},
show that the quantiles of $H^{w,2}$ and $H^{w}$ are effectively
equal for all the listed datasets, whereas the equivalent distributions
computed via clique projection are much narrower.

Next, the intermediate behaviour of Class~2
is only evident for the Plant Pollination~(015) network. This is evident
from the quantile-quantile distance relations reported in Fig.~\ref{lower_dominance},
which show that the distributions of quantiles converge for an intermediate
layer number~$k$, which corresponds to the smallest layer for which all
the nodes in the hypergraph LCC are connected.

Finally, the rest of the datasets demonstrate a vector metric structure
that depends non-trivially on higher-order interactions, and thus they
belong to Class~3. The quantile-quantile plots of their distance distributions,
illustrated in Fig.~\ref{no_dominance}, show that the quantiles keep changing
with the addition of new layers. This indicates that there is no single
order whose interactions have a dominating effect in determining the hypergraph
distances. In turn, this may be due to a multitude of factors. For example,
the size distribution of the hyperedges may be heavy-tailed, uniform, or
even increasing with size. Alternatively, the interplay of the fraction
of hyperedges of a given size and the distribution of the sizes of their
intersections may yield the same effect. Either way, for networks in this
class, one needs to account for all orders of interaction to ensure that
no information is lost.

Specifically, in the case of NDC-class data, layer $k=4$ is the first
layer in which almost all nodes in the LCC are connected. However, the
quantiles of the distance distributions in $H^{w,4}$ and in $H^w$ are
significantly different, as they are for $k=5$ and $k=6$. Qualitatively,
the same effect appears for the NDC-substance network and the Senate-committee
data. However, in these cases the deviation between distributions is
much smaller.

The case of Genes and diseases data is the most interesting,
as deviations appear only for the layer $k=4$. This means that
the inclusion of interactions of size~5 significantly changes
the distances between the node pairs that become connected on
layer~4. However, the pairs that become connected on layers~2
and~3 are not affected by the higher-order interactions and,
therefore, may be considered as a kind of core with lower-order
interaction dominance. A partially similar picture appears in
the Plant Pollinators~(062) case, where a pairwise-interaction
core still appears, but with a significant shift of the distributions
for layers~3, 4 and~5.

\section{Conclusions}
In summary, we generalized three measures that describe complex
network topology to the case of higher-order interactions. Specifically,
we considered closeness and betweenness centralities, and node
eccentricity. 
All these are metric observables, as they depend
on the distances between nodes. Thus, we studied the effect on
them of adopting the recently proposed generalization of hypergraph
distance that explicitly accounts for hyperedges weights, hyperedges
sizes and sizes of hyperedge intersections~\cite{vasilieva}.

In a series of tractable examples, which may be considered
basic hypergraph motifs, we showed the differences induced
by the use of hypergraph distance with respect to the other
main method of computing distances, namely clique projection.
Our analyses show that measures computed via hypergraph distance
preserve important information about the higher-order topology
of the networks. Similar conclusions are obtained on a series
of real-world hypergraphs of diverse nature, for which we
show that three general classes of topological structure exist.
In particular, networks in one class are dominated by pairwise
interactions, and therefore they are effectively equivalent
to their clique projection. Networks belonging to the second
class are dominated by interactions whose order is intermediate
between pairwise ones and those with the largest size of hyperedge.
Finally, networks in the third class have a nontrivial distribution
of hyperedge sizes and intersections, so that no particular
order of interaction is negligible with respect to the rest.

These considerations are corroborated by an analysis
based on the multilayer representation of the hypergraphs,
obtained by introducing a vector of distances where
each component corresponds to a progressive filtration
layer in the hypergraph. Thus, our results provide a
natural way to identify the most relevant orders of
interaction in higher-order networks, allowing one to
operate an informed choice of approximation in the representation
of complex systems. In turn, this can help significantly
reducing the complexity of analyzing such systems, while
retaining most or all the information encoded in the
higher-order interactions.

Some future directions are as follows. First, the present
framework applies to undirected hypergraphs; extending the hypergraph
distance and the derived centrality measures to directed higher-order
structures is a natural next step. Second, the computational cost of
computing exact hypergraph distances grows rapidly with network size;
scalable approximations will be necessary for large-scale applications.
Third, the three-class taxonomy introduced here could serve as a
practical preprocessing step for machine-learning tasks on higher-order
networks, selecting the appropriate representation complexity before
downstream analysis.
Moreover, we note that the work~\cite{vasilyeva2023distances} additionally
generalizes global network efficiency using the same hypergraph distance,
demonstrating further possibilities in dataset rankings (e.g.\ the Senate committees
network ranks as the least efficient structure under hypergraph distance, despite
appearing highly efficient under clique projection); incorporating this fourth
measure would further strengthen the present results.

\section*{Acknowledgements}
This work was supported by the Ministry of Economic Development of the Russian Federation (code 25-139-66879-1-0003).

\section*{Data availability}
The data sets used are freely available at
\begin{itemize}
 \item[{}] \url{https://www.cs.cornell.edu/~arb/data/}
 \item[{}] \url{https://xgi.readthedocs.io/en/stable/xgi-data.html}
 \item[{}] \url{https://sociopatterns.org}
 \item[{}] \url{https://github.com/mattbierbaum/arxiv-public-datasets}
\end{itemize}

\section*{Code Availability}
The source code for all the computations is available with the raw data of the results
at\\ \url{https://codeberg.org/paraw/Hyperdist2/}

\bibliographystyle{unsrt}

\begin{thebibliography}{10}
	
	\bibitem{newman2003structure}
	Mark Newman.
	\newblock The structure and function of complex networks.
	\newblock {\it SIAM Review}, 45:167, 2003.
	
	\bibitem{boccaletti2006complex}
	Stefano Boccaletti, Vito Latora, Yamir Moreno, Martin Chavez, and D-U Hwang.
	\newblock Complex networks: Structure and dynamics.
	\newblock {\it Physics Reports}, 424:175, 2006.
	
	\bibitem{estrada2012structure}
	Ernesto Estrada.
	\newblock {\it The structure of complex networks: theory and applications}.
	\newblock Oxford University Press, 2012.
	
	\bibitem{boccaletti2014multi}
	Stefano Boccaletti, Ginestra Bianconi, Regino Criado, Charo~Ivan del Genio,
	Jesús Gómez-Gardeñes, Miguel Romance, Irene Sendiña-Nadal, Zhen Wang, and
	Massimiliano Zanin.
	\newblock The structure and dynamics of multilayer networks.
	\newblock {\it Physics Reports}, 544:1, 2014.
	
	\bibitem{newman2018networks}
	Mark Newman.
	\newblock {\it Networks}.
	\newblock Oxford university press, 2018.
	
	\bibitem{battiston2020networks}
	Federico Battiston, Giulia Cencetti, Iacopo Iacopini, Vito Latora, Maxime
	Lucas, Alice Patania, Jean-Gabriel Young, and Giovanni Petri.
	\newblock Networks beyond pairwise interactions: Structure and dynamics.
	\newblock {\it Physics reports}, 874:1, 2020.
	
	\bibitem{boccaletti2023structure}
	Stefano Boccaletti, Pietro De~Lellis, Charo~Ivan del Genio, Karin
	Alfaro-Bittner, Regino Criado, Sarika Jalan, and Miguel Romance.
	\newblock The structure and dynamics of networks with higher order
	interactions.
	\newblock {\it Physics Reports}, 1018:1, 2023.
	
	\bibitem{recon25}
	Yin-Jie Ma, Zhi-Qiang Jiang, Fanshu Fang, Charo~I. del Genio, and Stefano
	Boccaletti.
	\newblock Reconstructing simplicial complexes from evolutionary games.
	\newblock {\it Physical Review E}, 111:044304, 2025.
	
	\bibitem{nearidsimp25}
	Fatemeh Parastesh, Mahtab Mehrabbeik, Karthikeyan Rajagopal, Sajad Jafari,
	Matjaž Perc, Charo~I. del Genio, and Stefano Boccaletti.
	\newblock Synchronization stability in simplicial complexes of near-identical
	systems.
	\newblock {\it Physical Review Research}, 7:033039, 2025.
	
	\bibitem{hypermod}
	Charo~I. del Genio.
	\newblock Hypermodularity and community detection in hypergraphs.
	\newblock {\it Physical Review Research}, 7:033045, 2025.
	
	\bibitem{solst25}
	Vladimir~V Semenov, Subhasanket Dutta, Stefano Boccaletti, Charo~I. del Genio,
	Sarika Jalan, and Anna Zakharova.
	\newblock Solitary states in spiking oscillators with higher-order
	interactions.
	\newblock {\it Physical Review E}, 112:034302, 2025.
	
	\bibitem{equi26}
	K~Kovalenko, G~Contreras-Aso, C~I del Genio, S~Boccaletti, and R~J
	Sánchez-García.
	\newblock Equitability and explosive synchronisation in multiplex and
	higher-order networks.
	\newblock {\it Communications Physics}, 9:117, 2026.
	
	\bibitem{highchim26}
	Z~Guo, Z~Lui, S~Guan, C~I del Genio, S~Boccaletti, and J~Zhou.
	\newblock Higher-order interactions induce chimera states in globally coupled
	oscillators.
	\newblock {\it Physical Review Research}, 8:023011, 2026.
	
	\bibitem{aksoy2020hypernetwork}
	Sinan~G Aksoy, Cliff Joslyn, Carlos~Ortiz Marrero, Brenda Praggastis, and
	Emilie Purvine.
	\newblock Hypernetwork science via high-order hypergraph walks.
	\newblock {\it EPJ Data Science}, 9:16, 2020.
	
	\bibitem{nortier2025higher}
	Berné~L Nortier, Simon Dobson, and Federico Battiston.
	\newblock Higher-order shortest paths in hypergraphs.
	\newblock {\it Physical Review E}, 112:054302, 2025.
	
	\bibitem{vasilieva}
	Charo~I del Genio, Ekaterina Vasilyeva, Liubov Tupikina, Dmitry Fedorov, Daniil
	Musatov, Andrei~M Raigorodskii, and Stefano Boccaletti.
	\newblock Distances in weighted higher-order networks.
	\newblock {\it Communications Physics}, 2026.
	
	\bibitem{vasilyeva2023distances}
	Ekaterina Vasilyeva, Miguel Romance, Ivan Samoylenko, Kirill Kovalenko, Daniil
	Musatov, Andrey~Mihailovich Raigorodskii, and Stefano Boccaletti.
	\newblock Distances in higher-order networks and the metric structure of
	hypergraphs.
	\newblock {\it Entropy}, 25:923, 2023.
	
	\bibitem{vasilyeva2021multilayer}
	E~Vasilyeva, A~Kozlov, Karin Alfaro-Bittner, D~Musatov, AM~Raigorodskii,
	Matjaž Perc, and Stefano Boccaletti.
	\newblock Multilayer representation of collaboration networks with higher-order
	interactions.
	\newblock {\it Scientific reports}, 11:5666, 2021.
	
	\bibitem{puzis2013betweenness}
	Rami Puzis, Manish Purohit, and VS~Subrahmanian.
	\newblock Betweenness computation in the single graph representation of
	hypergraphs.
	\newblock {\it Social networks}, 35:561, 2013.
	
	\bibitem{Benson-2018-simplicial}
	Austin~R. Benson, Rediet Abebe, Michael~T. Schaub, Ali Jadbabaie, and Jon
	Kleinberg.
	\newblock Simplicial closure and higher-order link prediction.
	\newblock {\it Proceedings of the National Academy of Sciences}, 115:E11221,
	2018.
	
	\bibitem{Mastrandrea-2015-contact}
	Rossana Mastrandrea, Rossana Fournet, and Alain Barrat.
	\newblock Contact patterns in a high school: A comparison between data
	collected using wearable sensors, contact diaries and friendship surveys.
	\newblock {\it PLoS One}, 10:e0136497, 2015.
	
	\bibitem{chodrow2021generative}
	Philip~S Chodrow, Nate Veldt, and Austin~R Benson.
	\newblock Generative hypergraph clustering: From blockmodels to modularity.
	\newblock {\it Science Advances}, 7:eabh1303, 2021.
	
	\bibitem{goh2007human}
	Kwang-Il Goh, Michael~E Cusick, David Valle, Barton Childs, Marc Vidal, and
	Albert-László Barabási.
	\newblock The human disease network.
	\newblock {\it Proceedings of the National Academy of Sciences}, 104:8685,
	2007.
	
	\bibitem{petanidou1993pollination}
	T~Petanidou and D~Vokou.
	\newblock Pollination ecology in a phryganic ecosystem.
	\newblock {\it American Journal of Botany}, 80:892, 1993.
	
	\bibitem{robertson1929flowers}
	Charles Robertson.
	\newblock {\it Flowers and insects: lists of visitors of four hundred and
		fifty-three flowers}.
	\newblock Science Press, 1929.
	
	\bibitem{isella2011s}
	Lorenzo Isella, Juliette Stehlé, Alain Barrat, Ciro Catuto, Jean-François
	Pinton, and Wouter Van~den Broeck.
	\newblock What's in a crowd? analysis of face-to-face behavioral networks.
	\newblock {\it Journal of theoretical biology}, 271:166, 2011.
	
	\bibitem{clement2019arxiv}
	Colin~B. Clement, Matthew Bierbaum, Kevin~P. O'Keeffe, and Alexander~A. Alemi.
	\newblock On the use of arxiv as a dataset.
	\newblock {\it arXiv}, page 1905.00075, 2019.
	
	\bibitem{kovalenko2022vector}
	Kirill Kovalenko, Miguel Romance, Ekaterina Vasilyeva, David Aleja, Regino
	Criado, Daniil Musatov, Andrei~M Raigorodskii, Julio Flores, Ivan Samoylenko,
	Karin Alfaro-Bittner, Matjaž Perc, and Stefano Boccaletti.
	\newblock Vector centrality in hypergraphs.
	\newblock {\it Chaos, Solitons \& Fractals}, 162:112397, 2022.
	
\end{thebibliography}

\end{document}